\documentclass[12pt,preprint]{aastex}
\usepackage{natbib}
\usepackage{amsfonts}
\usepackage{color}
\usepackage{ulem}
\usepackage{amsmath}
\usepackage{mathtools}
\usepackage{longtable,lscape}
\newcommand{\rsun}{$R_{S}\,$}

\newcommand{\eb}{\mathbf{\hat{e}_{b}}}
\newcommand{\mfp}{\ensuremath{\,{\lambda_{\parallel}}\,}}

\shorttitle{Modeling Coronal Proton Acceleration}
\shortauthors{K. A. Kozarev, R. M. Evans, N. A. Schwadron, M. A. Dayeh, M. Opher, K. E. Korreck, B. van der Holst}

\begin{document}

\title{Global Numerical Modeling of Energetic Proton Acceleration \\
	in a CME Traveling Through the Solar Corona}


\author{Kamen A. Kozarev\altaffilmark{1,5}, Rebekah M. Evans\altaffilmark{2}, Nathan A. Schwadron\altaffilmark{3}, Maher A. Dayeh\altaffilmark{4}, Merav Opher\altaffilmark{1}, Kelly E. Korreck\altaffilmark{5}, Bart van der Holst\altaffilmark{6}}

\altaffiltext{1}{Department of Astronomy, Boston University}
\altaffiltext{2}{NASA/Goddard Space Flight Center}
\altaffiltext{3}{Institute for the Study of Earth, Oceans, and Space, University of New Hampshire}
\altaffiltext{4}{Department of Space Science, Southwest Research Institute}
\altaffiltext{5}{Harvard-Smithsonian Center for Astrophysics}
\altaffiltext{6}{Center for Space Environment Modeling, University of Michigan}

%\affil{\\Astronomy Department,\\Boston University,\\
 %   Boston, MA 02134\\~\\~\\}
%\affil{\\NASA/Goddard Space Flight Center,\\
  %  Greenbelt, MD 20771\\~\\~\\}


%\newpage

%\pagenumbering{roman}
%\tableofcontents

%\tableofcontents
%\clearpage
%\pagenumbering{arabic}

\begin{abstract}

The acceleration of protons and electrons to high (sometimes GeV/nucleon) energies by solar phenomena is a key component of space weather. These solar energetic particle (SEP) events can damage spacecraft and communications, as well present radiation hazards to humans. In depth particle acceleration simulations have been performed for idealized magnetic fields for diffusive acceleration and particle propagation, and at the same time the quality of MHD simulations of CMEs has improved significantly. However, to date these two pieces of the same puzzle have remained largely decoupled. Such structures may contain not just a shock, but also sizable sheath and pile-up compression regions behind it, and may vary considerably with longitude and latitude based on the underlying coronal conditions. In this work, we have coupled results from a detailed global 3D MHD time-dependent CME simulation to a global proton acceleration and transport model, in order to study time-dependent effects  of SEP acceleration between 1.8 and 8 solar radii in the 2005 13 May CME. We find that the source population is accelerated to at least 100~MeV, with distributions enhanced up to six orders of magnitude. Acceleration efficiency varies strongly along field lines probing different regions of the dynamically evolving CME, whose dynamics is influenced by the large-scale coronal magnetic field structure. We observe strong acceleration in sheath regions immediately behind the shock.

%We further simulate the proton transport between the Sun and Earth, and find the modeled fluxes consistent with particle observations.

%REBEKAH'S ABSTRACT VERSION:
%The acceleration of protons and electrons to high (sometimes GeV/nucleon) energies by solar phenomena is a key component of space weather. These solar energetic particle (SEP) events can damage spacecraft and communications, as well present radiation hazards to humans. In depth particle acceleration simulations have been performed for idealized magnetic fields for diffusive acceleration and particle propagation, and at the same time the quality of MHD simulations of CMEs has improved significantly. However, to date these two pieces of the same puzzle have remained largely decoupled. In this work, we provide the first comprehensive study of proton acceleration in a time-dependent global fashion by coupling a realistic CME and solar wind global MHD model to a kinetic particle transport and acceleration solver. We simulate the 13 May 2005 Earth-directed CME, and inject particles based on BLAH BLAH. We find: (list major results). In any case, the finite size and orientation of the coronal streamer, through which this CME propagates, creates a spatial and temporal difference in the shock strength along its front. This should not be overlooked when studying particle acceleration in a realistic coronal setting. This tool has to capability to shed light on the acceleration of particles at CME-driven shocks in a way unlike any previous study. 

\end{abstract}

\clearpage
\newpage
%=============================================END=============================================

%=============================INTRODUCTION===================================
\section{Introduction}

%\begin{figure}
%\noindent\includegraphics[width=1.0\columnwidth]{figures/mann_alfven_profile.png}
%\caption{Alfven speed profile in the solar corona. The three lines represent small changes to the magnetic model used. Figure taken from \citet{Mann:2003}.}
%\label{figmann_alfven_profile}
%\end{figure}

It has been established that charged solar energetic particles (SEPs) can be produced by various dynamic processes in the solar corona during solar eruptions. Of those, the main two are considered to be acceleration in flares and coronal mass ejections (CMEs) (See a review in \citet{Aschwanden:2006}). Here, we focus on the latterÕs capability to energize protons to energies above 1 MeV. \citet{Gopalswamy:2005} studied the CME heights at times of particle releases, and found that acceleration began at heights from 1.4 to 8.7 solar radii. Similarly, \citet{Reames:2009a} and \citet{Reames:2009b} studied ion coronal release times and locations based on in-situ observations of different species at a variety of energies, and deduced that particles could be accelerated to SEP energies as low as 2 solar radii. This is in agreement with modeling work by \citet{Zank:2000}, who studied maximum energies to which protons could be accelerated in coronal shocks. More recently, remote observations have confirmed that CME-driven shock waves can occur low in the corona, which means they may also accelerate SEPs there \citep{Kozarev:2011, Ma:2011, Veronig:2010, Liu:2010}.

In depth simulations of particle-in-cell acceleration have been performed for idealized interplanetary magnetic fields to test theories of diffusive acceleration \citep{Giacalone:2005a}. Theoretical estimates based on numerical simulations of realistic CMEs and related shocks have shown that the diffusive shock acceleration (DSA) process is capable of accelerating solar energetic protons up to energies of 10 GeV for very strong shocks \citep{Roussev:2004, Manchester:2005, Kocharov:2005, Tsurutani:2003}. However, even as the quality of detailed MHD simulations of CMEs has improved significantly, they have remained largely decoupled from detailed numerical models of particle acceleration and propagation. Most global shock acceleration and particle propagation models focus on interplanetary space between $\sim$20 solar radii and 1 astronomical unit (AU) \citep{Verkhoglyadova:2010, Zhang:2009}, and neglect the region in the corona between 1.2 and 10 solar radii, shown to be a dynamic and very important environment for the creation of energetic particles.

There have been few modeling studies of the dynamics of CMEs low in the corona and their effects on proton acceleration, mostly because of the dynamic nature of the region, complex geometries, and lack of in situ observations. \citet{Ng:2008} modeled particle acceleration at an idealized coronal shock starting from 3.5 Rs. They were able to show that the shock of constant speed 2500 km/s could accelerate protons to above 300 MeV in 10 minutes. \citet{Vainio:2007} and \citet{Vainio:2008} studied the effects of self-generated turbulence on particle acceleration at coronal shocks interacting with analytical flux rope geometries. \citet{Kocharov:2011} and \citet{Kocharov:2012} presented a numerical model for coronal shock acceleration on expanding flux tubes of a simple semicircular geometry, and studied the effects of shock orientations, as well as possible escape sites for particles.

In order to obtain more credible results for the acceleration of coronal ions over the duration of SEP events, a more accurate description of the plasma dynamics during event onset is necessary.  To that end, studies were performed by \citet{Sokolov:2004}, \citet{Kota:2005}, and \citet{Sokolov:2009}, in which the authors coupled time-dependent 3D numerical CME simulation results in the corona to two different particle propagation models. Those authors found that under the conditions they imposed protons could be accelerated to hundreds of MeV energies. However, those studies were meant to introduce the particle models, rather than to present in depth analyses of particle acceleration at coronal heights. More in-depth studies of energetic particle behavior early in events using such advanced modeling tools are needed in order to answer important questions about where acceleration is strongest as a function of radial distance and position relative to the CME propagation direction, and what coronal plasma and shock parameters influence acceleration. In addition, such modeling capability is necessary to support observation analysis for upcoming missions such as Solar Probe Plus and Solar Orbiter.

According to current understanding, the efficiency of particle acceleration may vary considerably along a CME shock wave, depending on the angle between the shock normal and the incident magnetic field. Diffusive acceleration theory predicts that parallel shocks are inefficient accelerators of coronal ions due to the long time scales required to energize particles relative to the average time that CME shocks spend in the low and middle corona. It is possible, however, that local self-excited waves can enhance particle acceleration \citep{Lee:2005,Ng:2008}. On the other hand, perpendicular shocks associated with CMEs low in the corona are good candidates for accelerating charged particles to $>$MeV energies because the perpendicular geometries allow for fast acceleration timescales \citep{Giacalone:2006}. Recent simulations by \citet{Giacalone:2008, Guo:2010} have shown that the time-dependent orientation of MHD shocks with respect to the magnetic geometry can significantly influence particle fluxes. Thus, realistic global simulations of the coronal magnetic fields and plasma environment, traveling shocks, and energetic particles are needed to address the questions of how, where, and when significant SEP fluxes are produced, adequately and in detail.

We present a study exploring in detail the effects of proton acceleration in the three-dimensional (3D) corona during the eruptive event on 2005 May 13. The goal is to study how the evolution of a realistic CME and the shock it drives influence proton spectra, and how the amount of proton acceleration varies as a function of position around the CME front. The Energetic Particle Radiation Environment Module (EPREM) model is well equipped for this study, as its 3D Lagrangian grid may readily accommodate the complex CME dynamics. We have modified this model for the coronal environment. The version used here does not include perpendicular (cross-field) diffusion. The implicit assumption is that the proton gyroradii are sufficiently small near the Sun to neglect particle scattering perpendicular to the magnetic field as a significant transport effect in the corona. This is not necessarily true since large-scale magnetic fluctuations can enhance the perpendicular diffusion \citep{Giacalone:1999, Matthaeus:2003}, which in turn has been shown to be important for particle acceleration at quasi-perpendicular shocks \citep{Zank:2006}. Perpendicular diffusion will be addressed in future work. 

The Space Weather Modeling Framework (SWMF), an advanced global MHD model, is used to simulate the background solar wind and a CME. It has been shown to reproduce known signatures of solar eruptions in the corona and interplanetary space \citep{Manchester:2008}, and has recently been modified to include the physics of coronal heating through wave dissipation \citep{vanderHolst:2010,Evans:2012}.

This manuscript is organized as follows: Section \ref{modelingframework} describes the models used, as well as the coupling between them. Section \ref{modeledtransients} presents time-dependent properties of the modeled CME. In Section \ref{coronalmodeling} we present the parameters of the energetic particle transport model. Results of the modeling and a discussion are in Section \ref{coronalresults}. Finally, we summarize the work in Section \ref{coronalsummary}.
%=============================CORONAL THEORY INTRO==================================

%=============================MODELING FRAMEWORK==================================
%Section 2
\section{Modeling Framework}
\label{modelingframework}

% 2.1. BATSRUS description
\subsection{Coronal and Solar Wind Model}
\label{mhdmodel}
In order to model the complex environment of the solar corona, results from SWMF \citep{Toth:2012} have been used in this work. The three dimensional magnetohydrodynamic (MHD) code Block Adaptive Tree Solar-Wind Roe Upwind Scheme (BATSRUS, as it will be called throughout the rest of the paper) serves as the core of the SWMF. BATSRUS is highly parallelized and includes adaptive mesh refinement \citep{Powell:1999}. This coronal model includes the physics of Alfven waves as a source of momentum to accelerate the fast solar wind, and wave dissipation to heat the corona \citep{Evans:2012}. A flat spectrum of Alfven waves is specified at the inner boundary along open field lines, and the wave energy density evolves according to a frequency-averaged wave transport equation in the WKB approximation\citep{Sokolov:2009}. The simulation used in this paper is one of the first self-consistent studies of global CME evolution in a physics-based solar wind model \citep{Manchester:2012}. A detailed description of all of the model's features is beyond the scope of this paper. The reader is advised to follow the references in this section for more information.

The work presented here uses results from a simulation of the 2005 May 13 CME event. The simulation domain is initially constructed as a Sun-centered box of size $48\times48\times48$ \rsun with six levels of refinement (each differing by a factor of two). The initial solar magnetic field is calculated with the Potential Field Source Surface model \citep{Altschuler:1969} and a Michelson Doppler Imager synoptic magnetogram of CR 2029. The potential solution is allowed to evolve according to the ideal MHD equations. Adaptive mesh refinement resolves the heliospheric current sheet during convergence to steady state. After the steady-state solution is achieved, a high resolution box is placed in the path of the CME's propagation. The purpose of the box is to eliminate the influence of jumps in grid refinement on the CME's evolution and capture the shock and ICME properties well near the nose. 

To initiate the eruption, a modified Titov-Demoulin (TD) flux rope \citep{Titov:1999, Roussev:2004} is inserted out of equilibrium in an active region. As it is unstable, it erupts due to instability and magnetic forces. For the modeling results presented here, the TD flux rope has been configured using only a line current running through the torus, and thus the ejecta field is poloidal. Observations of the prominence material and height-time measurements of the CME in the corona were used to constrain the flux rope parameters.

From the simulation results, a series of 3D boxes of size $13\times16\times16$\rsun ($141\times173\times173$ cells) around the expanding CME has been extracted at two-minute cadence. The two-minute cadence is sufficient in order to keep the particle transport calculations coupled to the changes in the plasma. The boxes are regularly gridded, and contain the instantaneous value of coronal/solar wind density, velocity, magnetic field, and thermal pressure at every cell. The cell size is 0.092 Rs.

\subsection{Particle Acceleration and Propagation Model}
\label{particlemodel}
% 2.2. EPREM description
The EPREM model \citep{Schwadron:2010} is a parallelized energetic particle transport and acceleration numerical kinetic code, solving for charged particle transport in the 3D heliosphere. Transport effects include particle streaming, convection with the solar wind flow, pitch-angle scattering, and adiabatic energy change. It uses a dynamical simulation grid, in which the computational nodes are carried away from the Sun (frozen-in) with the solar wind - thus the connected grid nodes (streamlines) naturally assume the shape of a three-dimensional interplanetary magnetic field, along which energetic particles propagate. The model has been validated for the interplanetary SEP transport \citep{Dayeh:2010, Kozarev:2010}. A slightly modified form of the focused transport equation \citep{Skilling:1971, Ruffolo:1995, Tylka:2001, Ng:2003} is used to treat transport and energy change, with coefficients specified in a way that they can be computed along nodes moving with the solar wind flow \citep{Kota:2005, Schwadron:2010}:

\begin{eqnarray}
\bigg(1 & - &\left. \frac{\mathbf{V}\cdot \eb v \mu}{c^2}
\right)\frac{df}{dt} \text{~~~~~~~~~~~~~~~~~~~~~~~~~~~~~~~~~~~~~~~~~~~~~~~~~~~~~~~~~~~~~~~~~~~~~~~~~(convection)}\nonumber\\
&+& v\mu \eb\cdot \nabla f \text{~~~~~~~~~~~~~~~~~~~~~~~~~~~~~~~~~~~~~~~~~~~~~~~~~~~~~~~~~~~~~~~~~~~~~~~~~~~~~~~(streaming)}\nonumber\\ 
& + & \frac{(1-\mu^2)}{2}\left[ v \eb\cdot \nabla \ln B -
  \frac{2}{v}\eb\cdot \frac{d\mathbf{V}}{dt} + \mu
  \frac{d\ln(n^2/B^3)}{dt} \right] \frac{\partial f}{\partial \mu } 
\nonumber \text{~~~~~~~~~~~~~~~~~~~~~~(focusing)}\nonumber\\ 
& + & \left[-\frac{\mu \eb}{v}\cdot \frac{d\mathbf{V}}{dt} +
  \mu^2 \frac{d\ln (n/B)}{dt} + \frac{(1-\mu^2)}{2}\frac{d\ln
 B}{dt}\right] \frac{\partial f}{\partial \ln p} \text{~~~~~~~~~~~~~~(adiabatic change)}\nonumber\\ 
 & = &\frac{\partial }{\partial \mu} \left( \frac{D_{\mu \mu}}{2}
\frac{\partial f}{\partial \mu}\right) \text{~~~~~~~~~~~~~~~~~~~~~~~~~~~~~~~~~~~~~~~~~~~~~~~~~~~~~~~~~(pitch-angle scattering)}\nonumber\\ 
%&-& \frac{1}{p^2}\frac{\partial}{\partial p} \left( p^2 D_{pp} \frac{\partial f_0}{\partial p}\right) \text{~~~~~~~~~~~~~~~~~~~~~~~~~~~~~~~~~~~~~~~~~~~~~~~(stochastic acceleration)}\nonumber\\ 
&+& Q.
\label{eq:focusedTransport}
\end{eqnarray}
Here $\eb$ is the unit vector along the magnetic field, $\mu$ is the cosine of the particle pitch-angle, $n$ is the solar wind number density, $B$ is the magnetic field strength, $p$ is momentum,  $Q$ is a source term, with a pitch-angle diffusion coefficient
\begin{eqnarray}
%D_{\mu\mu} & = & \left( \frac{R_1}{r}\right)^{3/2}\frac{(1-\mu^2)v}{2\lambda_0},
D_{\mu\mu} = \frac{(1-\mu^2)v}{2\lambda_{||}}.
\label{eq:Dmumu}
\end{eqnarray}
where the parallel mean free path is $\lambda_{||}$.
%where the parallel mean free path at $R_1 = 1$ AU is $\lambda_0$., and the coefficient associated with diffusion in particle momentum is 
%
%\begin{eqnarray}
%\frac{D_{pp}}{p^2} = \eta^2 D_0 w
%\label{eq:Dpp}
%\end{eqnarray}
%where the ratio of the particle speed to the solar wind speed is $w = v/V$, the ensemble averaged square of field variations is $\eta^2 = \langle (B - B_0)^2/B_0^2\rangle$, $B_0$ is the mean magnetic field, and a constant $D_0 = 4\times 10^{-6} $ s$^{-1}$ is used as the rate of stochastic acceleration \citep{Schwadron:1996}. An average value of $\eta^2 = 0.07$, established from Ulysses spacecraft observations, is used here \citep{Schwadron:1996}. 
The most important terms for particle acceleration are the first-order adiabatic energy change terms above. They correspond to compressive acceleration in shocks and compression regions, or adiabatic cooling in interplanetary space. 

The modular design of EPREM allows for the underlying model for the interplanetary magnetic field to be changed easily. In its original formulation, EPREM employs the so-called Parker spiral: a radial field component falling off as the inverse square of radial distance, azimuthal component falling off as the inverse of radial distance, and a constant latitudinal component. The model treats the acceleration and transport of a distribution of protons with a dependence on pitch angle, momentum, time, and position. Along each field line (a connected list of nodes) the model solves for particle transport, adiabatic focusing and energy change, convection, and pitch-angle scattering.

\subsection{Coupling Between EPREM and BATSRUS}
\label{eprembatsrus}
The BATSRUS model has been shown to simulate realistically both steady state solar wind and time-dependent effects of propagating CMEs. Therefore, selecting the BATSRUS simulation data in the EPREM model is a significant improvement over the Parker spiral solar wind description. This change allows us to explore more details of SEP acceleration in the solar corona.

The kinetic model has been coupled to the results of time-dependent 3D MHD simulations by tri-linear interpolation of the EPREM nodes onto the MHD grid. This scheme takes into account the distances to cell edges in all three dimensions simultaneously, and is well suited to obtain interpolated values, since the interpolated boxes from BATSRUS are on a regular grid with a sufficiently small cell size.

Three-dimensional boxes have been extracted around the CME region. The time dependent 3D MHD boxes have been specially constructed to maintain cell positions constant over time. This can be used to smooth the behavior of the coronal plasma in time, so that there are no sharp jumps in the MHD quantities on timescales greater than the EPREM macro time step (which for this application has been set at 0.1 times the light crossing time between the Sun and Earth, or about 50 seconds). The coupling interface performs linear interpolation in time to ensure the smooth transition of the solar wind parameters in the EPREM grid between consecutive time steps of the MHD simulations. In practice this results in one or two extra steps between two consecutive MHD boxes. 

%Figure \ref{epremgrid} shows the result of coupling the MHD results from BATSRUS into the EPREM grid, for a single snapshot of a coronal simulation run. The EPREM nodes are shown in red with white lines connecting nodes along field lines. The blue frame denotes the boundaries of the MHD box. The MHD boxes only partially overlap with the EPREM grid, where the CME propagates.
% \begin{figure}
 %\noindent\includegraphics[width=1.0\columnwidth]{figures/epremgrid.eps}
 %\caption[EPREM grid for the coronal simulation]{A depiction of the EPREM grid nodes at a single snapshot, showing how the time-dependent solar wind flow (inside the blue box frame) distorts the Lagrangian grid.}
 %\label{epremgrid}
% \end{figure}
%=============================MODELING FRAMEWORK==================================

%===================================PLASMA STRUCTURES PROPERTIES===================================
\section{Properties of the Modeled Coronal Transient Structures}
\label{modeledtransients}

%1. OBEDINI INNER AND OUTER REGION PLASMA PROPERTIES V EDNA SEKCIQ!
%2. SYSHTOTO ZA INNER AND OUTER PARTICLE PROPERTIES!
%4. Da ima edna figura, 4 snapshota, za vseki mhd parameter... Zavisi dali shte puskame v ApJ ili ApjL!
%3. AKO OSTANE VREME, NAPRAVI GRAFIKA NA ACCELERATION RATE VS. RADIAL DISTANCE!

\subsection{Proposed Particle Acceleration Sites in CMEs}

\citet{Roussev:2004} explored the potential of a coronal shock for accelerating protons. They suggest that the simulated CME-driven shock can easily account for prompt energetic particle production below 10 GeV, mainly due to the high compression ratio it develops. However, their numbers are estimates based on DSA theoretical results.

Apart from the acceleration at the CME-driven shock, several other possible sites for the acceleration of charged particles have been discussed. \citet{Manchester:2005} perform an in-depth study of the behavior of the plasma in the sheath of a simulated fast CME propagating in an inhomogeneous solar wind. They find that acceleration of electrons is possible in a dimple on near-equatorial field lines draping around the CME. There, the magnetic field lines bend equatorward due to the nature of the shock, and then bend poleward in order to move around the flux rope. In addition, they find that strong velocity shears in the CME sheath could cause Kelvin-Helmholtz instabilities, which in turn may increase turbulence and the rate of particle acceleration. \citet{Kota:2005} also mention the points of great deflection of the magnetic field in low-latitude field lines as a very likely site of acceleration. They note that this deflection structure results from the CME pushing field lines out of its way near the Sun, and comment that the effect is likely to decay later on as the CME travels further out.

In recent work, \citet{Das:2011} explored the plasma structures that the model CMEs created during their propagation through the inner corona between 2 and 7 \rsun in two BATSRUS simulations. The two CMEs were driven by two models of out-of-equilibrium flux ropes inserted low in the corona. They noted the presence of a distinct feature in their simulations, which shows up as substantial increase in the plasma density and magnetic field. This piled up compression (PUC) region occurs in the sheath between the shock and the flux rope. In the work of \citet{Das:2011}, it is considered a result of the lateral overexpansion of the CME. The gradients in density between the PUC and the rest of the sheath are usually much larger than the density jump in the sheath behind the shock. The jump in density depends on the shape and amount of overexpansion of the CME. As will be seen below, it is not necessarily present at all times or around the entire CME, which is to be expected, given the uneven shape of the CME-pause in realistic simulations \citep{Evans:2011}. \citet{Das:2011} observed the PUC in both modeled CMEs, and it has also been observed in spacecraft measurements \citep{Dasso:2007}. \citet{Farrugia:2008} observe the pile up as a feature immediately preceding the flux rope at 1~AU, and also find lower lateral deflection speeds in that region, compared with the lateral speeds in the CME sheath. In effect, plasma structures may help accelerate coronal suprathermal protons to MeV energies.

In order to identify features of the CME, we use the angle $\theta_{Bz}=\sin^{-1}(B_Z/|\mathbf{B}|)$. It quantifies how much of the magnetic field points in the Z-direction. It has been used previously by \citet{Evans:2011} in a slightly different form to separate regions of different magnetic field source: sharp changes in this angle mark boundaries between different structures. A large gradient in this angle shows the interface between the CME flux rope and the solar wind field lines draped around it in the PUC region. In this work, we use the local minimum of this angle inward of the shock to identify a `central' location of the PUC.

%The defining characteristics of the plasma depletion layer(PDL) are a sharp jump and rotation of the magnetic field, and a minor dip in the density. The PDL was seen as a relatively narrow region sandwiched between the CME flux rope and the PUC in the work of \citet{Das:2011}. They hypothesized the PDL as the piling of solar wind magnetic field lines draping around the CME flux rope. The plasma on those field lines is then squeezed out along the field lines in keeping the total pressure balance. This phenomenon was described in connection with the magnetospheric bow shock by \citet{Zwan:1976}, and has been observed in the magnetosheath \citep{Farrugia:1997} as well as in CME sheaths \citep{Liu:2006}.

\subsection{Plasma Properties}
\label{plasmaproperties}
 
\begin{figure}
 \centering
 \noindent\includegraphics[width=0.8\columnwidth]{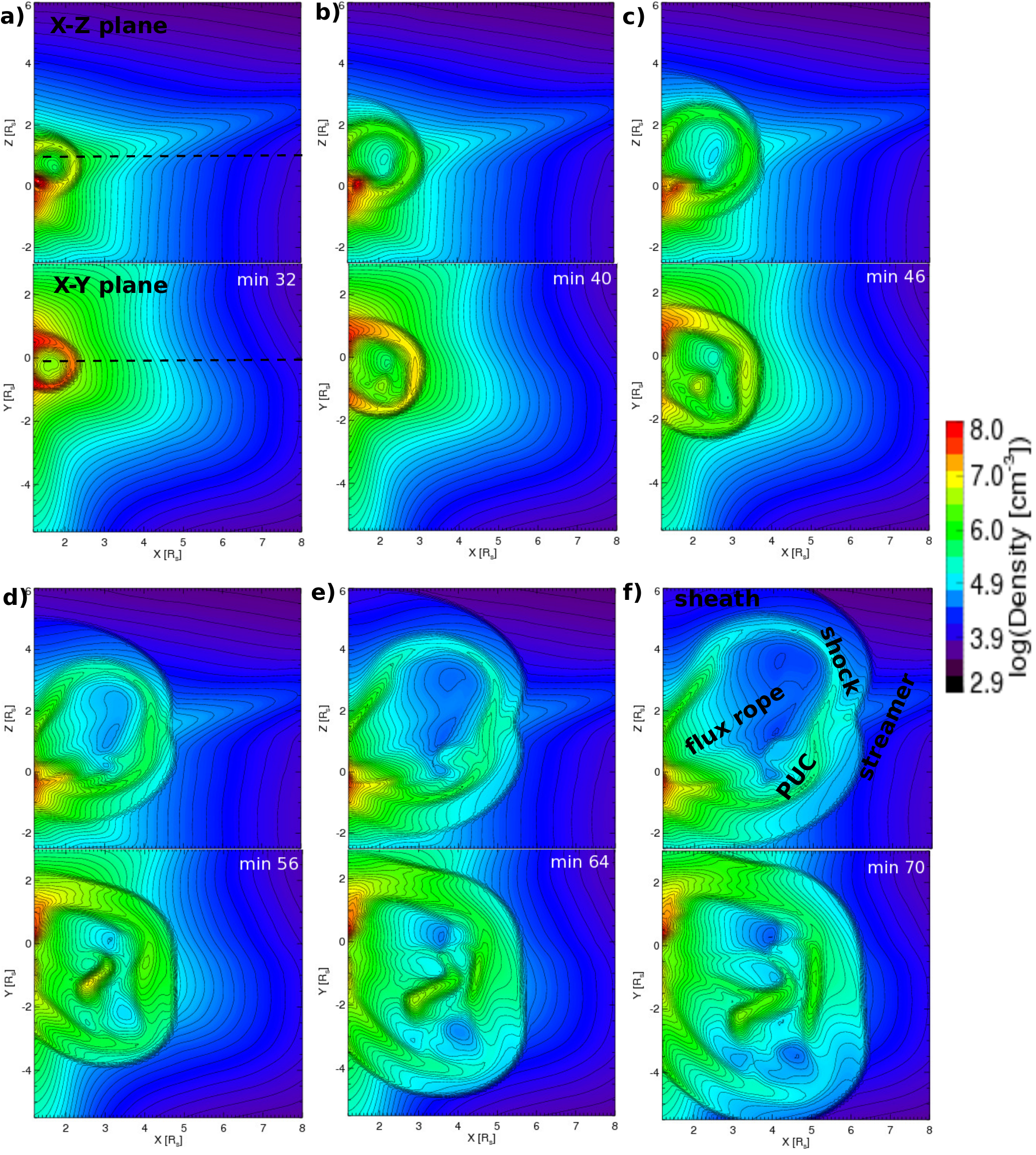}
 \caption{X-Z and X-Y slices showing color and line contours of proton density for six snapshots of the BATSRUS simulation. The snapshots span about 38 minutes of simulation time. The larger structure is a coronal streamer.}
 \label{figdensity}
 \end{figure}

Figure \ref{figdensity} shows the evolution of mass density at six snapshots in time, each represented by a double panel labeled a) through f). In every panel there are two slices showing density in the X-Z plane (top of panel) and in the X-Y plane (bottom of panel). The cuts were made at the Z=1 and Y=0 planes, and their locations are shown with horizontal dashed lines in a). The Sun is to the left of the panels. Time is measured relative to the same fiducial mark here as in all further discussion of this simulation.
The CME front travels between 2.0 and 6.5 \rsun in about 40 minutes, which gives an average speed of about 1300 km/s, in agreement with the CME speed inferred from remote sensing and in situ measurements \citep{Bisi:2010}. The line contours show density as well, to more clearly illustrate the changes. Concentrations of contour lines denote gradients in the density. The large scale time-independent structure of higher density, in which the CME propagates (large green structure in panel a), is a coronal streamer.

In the X-Z views, the CME resembles an elongated ellipsoid, which becomes distorted and develops a cavity as the front expands. The ellipsoid shape is due to the prescribed initial flux rope geometry. It retains a similar north-south orientation due to the relatively uniform density inside the streamer. After its northern edge reaches the edge of the high density streamer region, it begins overexpanding relative to the overall CME shape. In the last three snapshots the PUC is clearly seen separating from the rest of the CME sheath in the X-Z views. The various structures described here are denoted in panel f) as well.

In the X-Y views, the CME expands preferentially in the negative Y-direction, especially after it reaches the edge of the streamer. The sheath width increases over time, as well as density at the PUC. We only observe a clear distinction between the sheath and PUC in the last three cuts. In the central part of the CME some evolution of the flux rope is visible.

As the CME expands, it loses its initial symmetric shape. It expands preferentially to the north (positive Z) and east (negative Y). The shock and ejecta become more deformed due to the different solar wind density gradients at different latitudes as the CME expands beyond the coronal streamer above and below its densest portions. \citet{Manchester:2005} observed a similar ``localized inward deformation", or ``dimple", in the equatorial section of the shock, and proposed it as a possible site of particle acceleration due to the great deflection of the magnetic field there. The low density region behind the PUC is where the expanding flux rope is. As the flux rope and PUC become more stretched and irregular, the distance between them and the shock increases. By the time the CME reaches $\sim$7 \rsun, the sheath and the PUC have decoupled completely, with the latter remaining close to the flux rope, and the former occupying the rest of the space out to the shock. The PUC becomes relatively thinner as the sheath expands. Overall, the CME is most dense in the sheath (and PUC, where present) between the flux rope and the shock. 

\begin{figure}
\centering
\noindent\includegraphics[width=1.0\columnwidth]{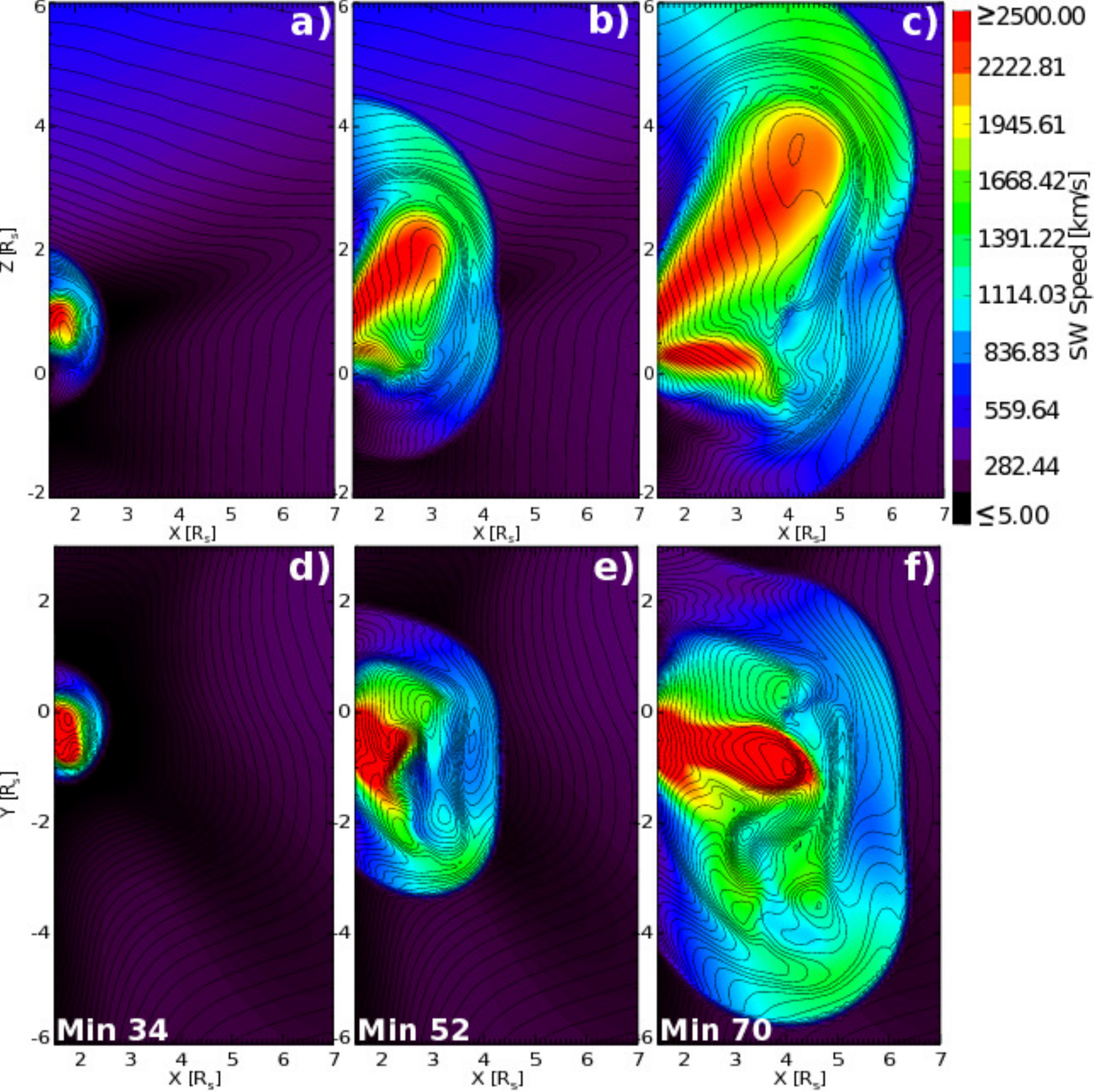}
\caption{Solar wind speed on two planar cuts for three times of the CME simulation. Panels a), b), and c) show the solar wind speed in an X-Z planar cut at minutes 34, 52, and 70. Panels d), e), and f) show the X-Y cuts at the same times. Density line contours are also plotted.}
\label{figusw}
\end{figure}

\begin{figure}
\centering
\noindent\includegraphics[width=1.0\columnwidth]{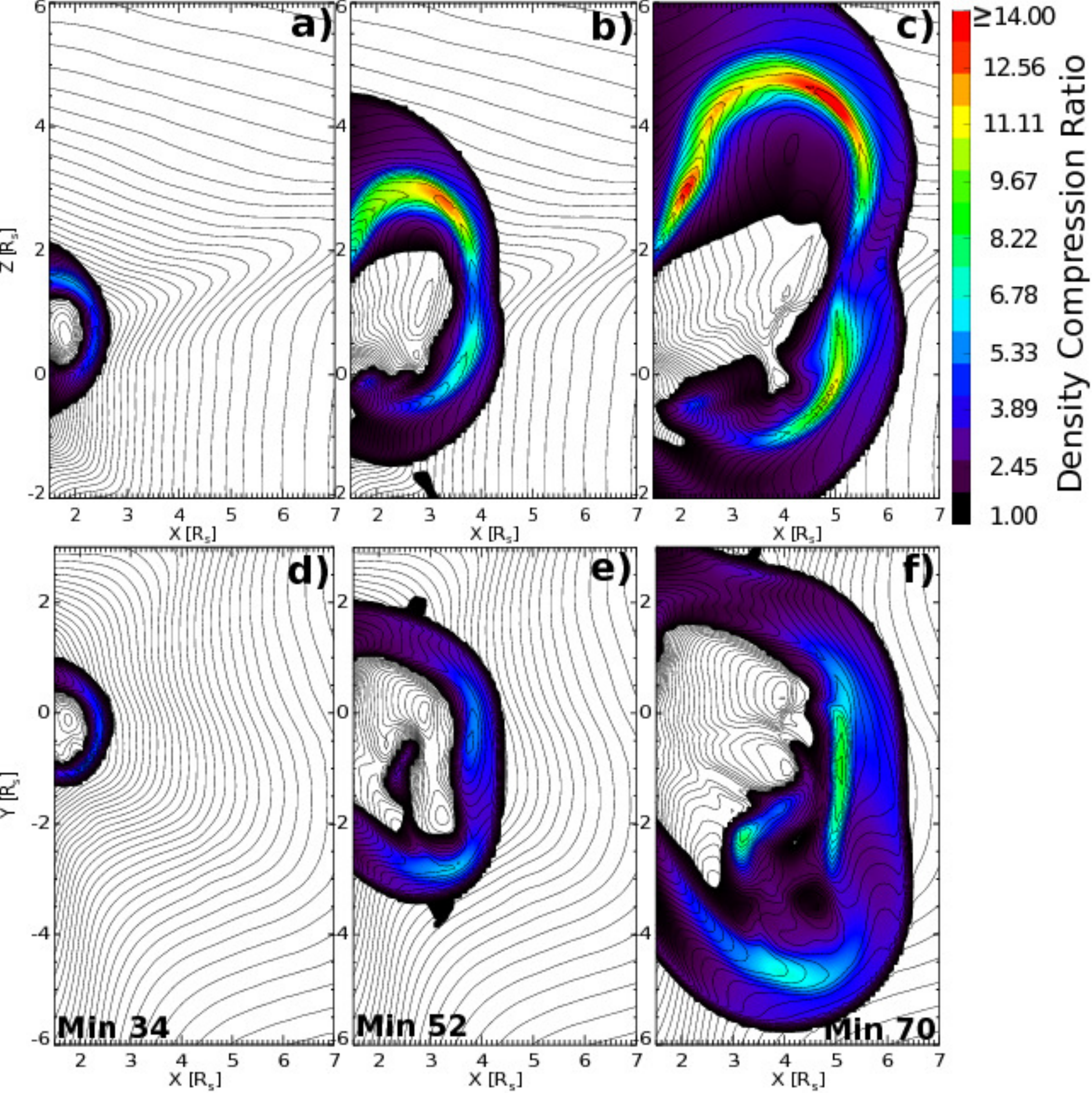}
\caption{Solar wind density compression ratios on two planar cuts for three times of the CME simulation. Panels a), b), and c) show compression ratios in an X-Z planar cut at minutes 34, 52, and 70. Panels d), e), and f) show the X-Y cuts at the same times. Density line contours are also plotted.}
\label{figcomp}
\end{figure}
 
The solar wind speed and density compression ratios are important parameters for the particle acceleration. Figures \ref{figusw} and \ref{figcomp} show the speed and compression ratio at the same X-Z (panels a)-c)) and X-Y (panels d)-f)) plane cuts as in \ref{figdensity}, for minutes 34, 52, and 70 of the CME simulation. The compression ratios are obtained by dividing the density values to the pre-eruption densities. This time ratio technique has been used previously in \citet{Das:2011} and \citet{Pomoell:2011}. We use it here to give a global view of the density compressions in the CME. Line contours of the density are overplotted in both figures. In Fig. \ref{figusw}, it can be seen that the solar wind speed is highest in the fast expanding CME flux rope, where it exceeds 2000~km/s throughout the simulation. The next highest speed region is the one discussed in the previous paragraphs - the overexpanding region in the sheath at high positive Z-values and high negative Y-values. It shows speeds between about 1200 and 1900~km/s. Comparing the speeds at the first and third times of the simulation, it can be seen that early on the expansion is more or less uniform, but at later stages there are large differences among the different regions of the CME. Fig. \ref{figcomp} shows that the highest density compression ratios occur consistently in the sheath in front of the CME flux rope, in a relatively thin region wrapped around the core of the CME. The highest ratios in that region exceed 12, compared with 1.5-4 in the rest of the sheath behind the shock. The ratios between the post-shock compressed densities and the PUC densities are thus between about 2 and 4.

The dynamics of the plasma transient presented here illustrate a general dependence of the CME dynamics on the particular conditions of the corona in which it expands. For example, had there not been a coronal streamer in its way, the CME would have propagated faster in all directions, and may have driven a stronger shock. In any case, the finite size and orientation of the coronal streamer, through which this CME propagates, creates a spatial and temporal difference in the shock strength along its front, as well as the plasma features following it. This should not be overlooked when studying particle acceleration in a realistic coronal setting.

%===============================END PLASMA STRUCTURES PROPERTIES===============================

%===================================CORONAL MODELING============================================
\section{Modeling Coronal Proton Acceleration with EPREM and BATSRUS}
\label{coronalmodeling}

%-------------------------------------------------------SUPRATHERMAL SOURCE----------------------------------------------------------
\subsection{Suprathermal Source at the Sun}
\label{suprathermalsource}
 \begin{figure}
 \noindent\includegraphics[width=1.0\columnwidth]{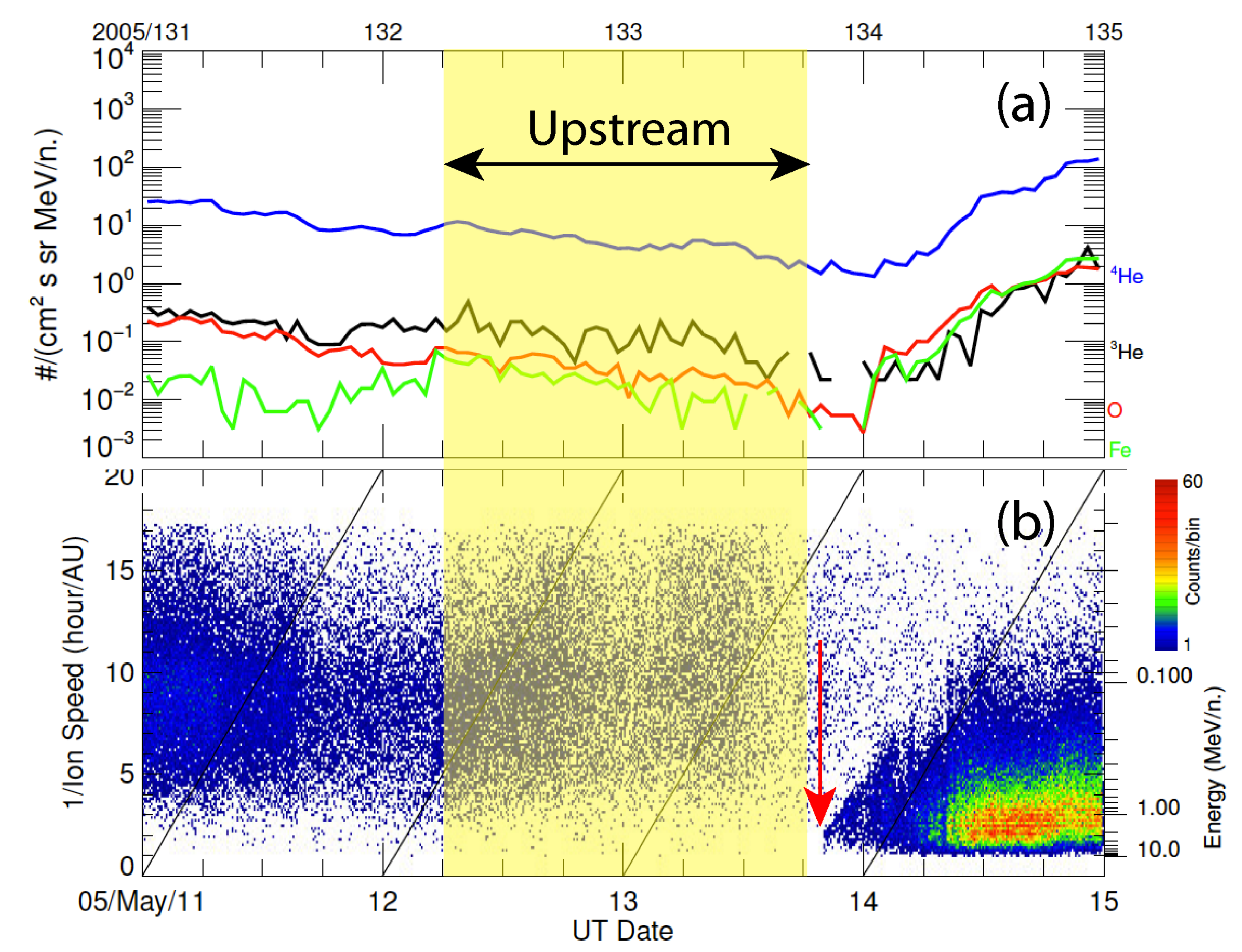}
 \caption{Intensity-time profiles of $\sim0.23-0.32$~MeV/nucleon, hourly averaged 3He, 4He, O, and Fe nuclei. (b) Energy spectrogram of $\sim0.03-10.0~MeV$ nucleon Fe-group ions. SEP event onset is indicated by the red vertical arrow and the yellow shaded region indicates the upstream sampling interval.}
 \label{figsuprathermal_observations}
 \end{figure}
\citet{Dayeh:2009} reported on measurements of suprathermal ions at 1 AU during carefully selected quiet times for most years of cycle 23. They found that the heavy ion abundances resembled the composition observed in corotating interaction regions (CIR) and unperturbed solar wind during solar minimum, while near solar maximum the heavy ion composition resembled that of large SEP events. This finding confirmed previous results that the suprathermal population within 1~AU is directly related to the remnant ions from solar activity \citep{Desai:2006}. Thus, the suprathermal particles observed at 1 AU likely reflect the source population near the Sun. Since these are quiet time abundances, it may be assumed that they are not modified significantly during their transit to 1~AU, and that the major transport effect on them is convection with the solar wind (and perhaps some adiabatic cooling) -  assuming also that they are emitted more or less isotropically from the solar surface.

%Here, we adopt a slightly different interpretation - that the suprathermal particles during the quiet times of solar maximum periods could result from the small scale flaring activity and magnetic reorganization of solar active regions, rather than from SEP events only. Nonetheless, the suprathermal particles observed at 1 AU still reflect the source population created at the Sun.

% are remnant SEP particles that are left over after major events and scatter over wide longitudinal and radial distances. A different interpretation, and the one supported here, is that those particles are a result of the increased activity on the Sun near solar maximum, when there is constant flaring and magnetic reorganization of the active regions. In a way, the suprathermal particles observed at 1 AU are very closely related to the source population created at the Sun. 
 
We used quiet-time observations of He$^{4}$ suprathermal (0.1-0.45~MeV/nuc) ions from ACE/ULEIS spacecraft over the 1.5-day quiet period before the 2005 May 13 event to calculate an average quiet-time source spectrum (Figure \ref{figsuprathermal_observations}) at the Sun. We scaled radially the observed spectrum, of the form dJ/dE=$0.52\pm0.18\times E^{-1.96\pm0.15}$ protons cm$^{-2}$ sr$^{-1}$ MeV$^{-1}$, and used it to represent the suprathermal source of SEPs near the Sun (The quiet spectrum over 0.5 days prior to the event onset was slightly quieter, but the spectral slope did not change, while the normalizing flux (J0) dropped by $\sim30\%$). We converted the Helium spectrum to a proton spectrum at 1.8 R$_{s}$ assuming: 1) the flux is scaled to 1.8 R$_{s}$ via an $r^{-2}$ dependence; 2) a helium to proton abundance of 10\% \citep{Lario:2003b}; 3) the resulting spectral form is valid between 0.1-1.0~MeV. After we applied these corrections, a pre-event spectrum of the form dJ/dE=$1.06\times 10^{5} \times E^{-1.96}$ protons cm$^{-2}$ sr$^{-1}$ MeV$^{-1}$ emerges. Beyond 1.0~MeV, we prescribed an exponential roll-over of the spectra, representing a lack of more energetic particles at the Sun during quiet times.

%---------------------------------------------------END SUPRATHERMAL SOURCE-------------------------------------------------------

%-------------------------------------------------------EPREM RUN PARAMETERS---------------------------------------------------------
\subsection{Coronal Simulation Parameters}
\label{coronalepremparameters}

The inner boundary for the EPREM simulation is at 1.8\rsun. The outer boundary varies for the individual field lines due to the varying dynamic conditions, but is in the range 10-13 \rsun, depending on the solar wind flow speeds. The computational grid has been set to be dense enough to resolve all features in the MHD results. The energy range of the injected proton spectra is 0.1-100 MeV, chosen to cover much of the SEP energy range relevant for space weather. The parallel mean free path has been assumed to equal 0.05 AU at 1 AU and 1 GV proton rigidity. It is additionally scaled with proton rigidity and radial distance from the Sun to reflect the magnetic turbulence spectrum and its radial dependence \citep{Zank:1998, Li:2003,Sokolov:2004,Verkhoglyadova:2009}:
\begin{equation}
\lambda_{||}= \lambda_{0} \left(\frac{pc}{1 GeV}\right)^{1/3} \left(\frac{R}{1 AU}\right)^{2/3},
\label{eqmfp_vs_r}
\end{equation}
where $\lambda_{0}=0.05$~AU. \citet{Sokolov:2004} used a similar treatment of \mfp; however they assumed a decaying upstream turbulence, which is hard to justify due to the lack of observations. \cite{Pomoell:2011} used a mean free path a hundred to a thousand times the particle gyroradii in the corona, depending on the particle rigidity. Here, the ratio ranges between $\sim900-50000$ for rigidities 0.1-100 MV. The value used here is similar to those in previous papers on proton transport \citep{Qin:2002, Zhang:2009, Kozarev:2010}. The values we use are similar to those commonly used. 

%In addition, a constant ratio of magnetic irregularities to the mean field is assumed for the stochastic acceleration in the focused transport equation: $\left(\frac{dB}{B}\right)^2=0.07$. In future work, we hope to put physical constraints on the diffusion coefficients using solar observations.

The scaled spectrum is injected uniformly at the inner boundary of the EPREM grid, without any time dependence. The steady state solution of the BATSRUS simulation (before the flux rope is inserted) is used to describe the solar corona during the initialization stage. The kinetic simulation is run for 50 minutes of simulation data with no time dependence, to ensure that steady state fluxes are achieved throughout the grid and for all energies. This corresponds to the quiet-time suprathermal seed particle spectra propagating to the outer boundary of the coronal model. 

After steady state is achieved, the simulation progresses in time by loading consecutive time states of the MHD solution, and computes the proton acceleration and transport based on the plasma conditions, while keeping the proton input spectrum constant on the inner boundary. The time steps are separated by roughly two minutes. Since the kinetic model's timestep is smaller than that, there is usually one or two time-interpolated time steps, where the MHD values are interpolated onto the EPREM grid using trilinear interpolation in space and linear interpolation in time. The simulation is run for 72 more minutes, until the large-scale shock front reaches $\sim$8 \rsun. The total simulation time is 122 minutes. The proton distribution function values are saved at every node of the grid, at every timestep and for each energy and pitch angle value. We extract solutions along field lines of interest and observe the response of the distribution function values to the evolving CME structures as a function of time.
%-------------------------------------------------------END EPREM RUN PARAMETERS---------------------------------------------------------

\section{Results and Discussion}
\label{coronalresults}

\begin{figure}%[h]
\centering
 \noindent\includegraphics[width=1.0\columnwidth]{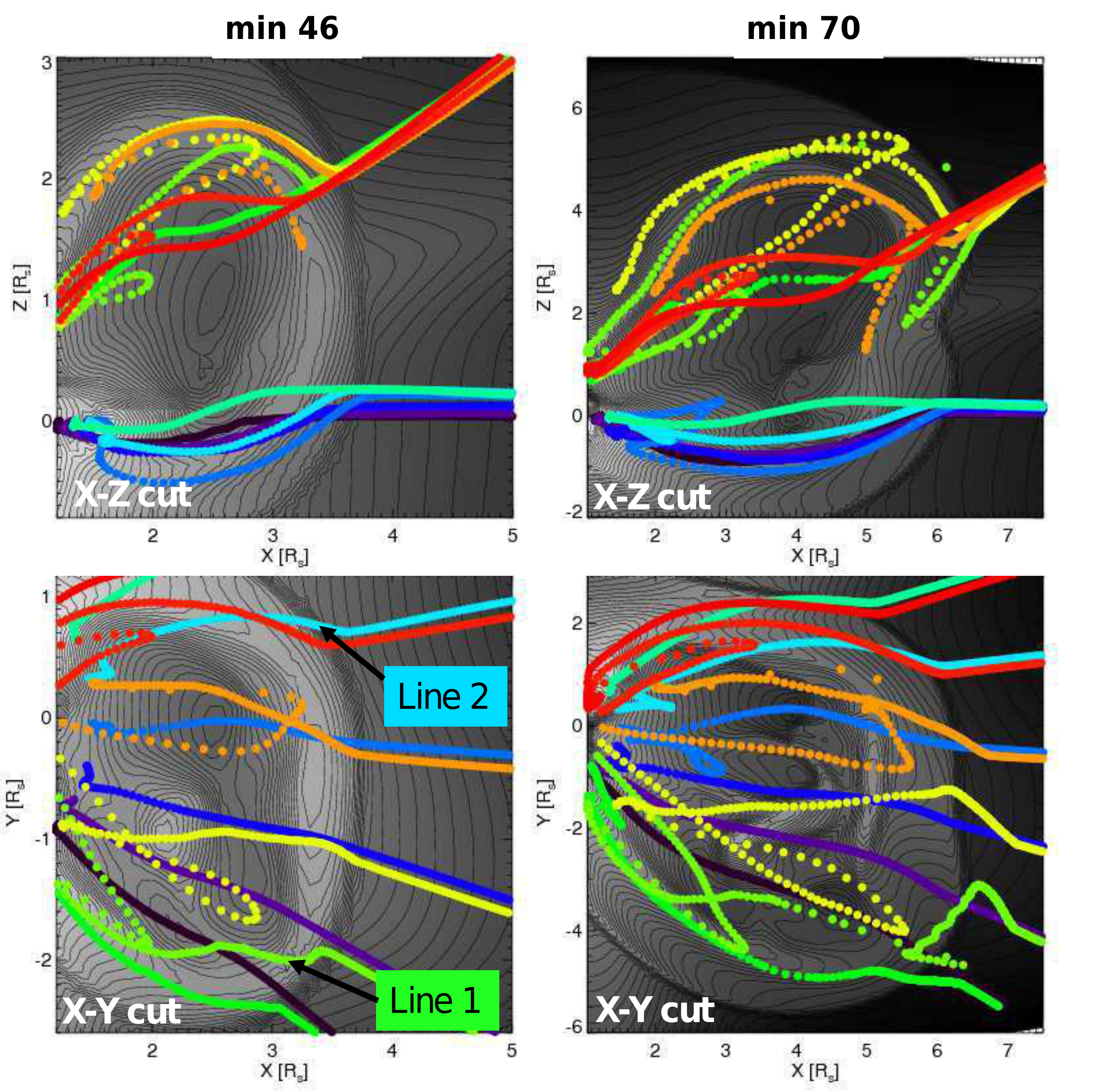}
 \caption{The evolving CME distorts the EPREM grid lines, as can be seen in these panels. Solar wind density is shown as grayscale density contours. Overlaid, color dots represent the grid node positions along individual field lines. The top and bottom panels show the X-Z and X-Y cuts for two times, similar to previous figures. The two lines marked `Line 1' and `Line 2' point to the two lines discussed later in this section.}
 \label{figeprem_color_lines}
 \end{figure}

We first look at the spatial distribution of the EPREM field lines within the MHD solution. Figure \ref{figeprem_color_lines} shows the evolution of the CME and the distortion it causes to the EPREM grid. In order to illustrate the effects that the CME/shock structure has on the energetic protons, we explore the acceleration and transport along two field lines. Of the 10 lines shown in Fig. \ref{figeprem_color_lines}, we select and compare two lines, in to demonstrate the extremes of particle behavior for different conditions in the simulation. Line 1 is located in the northeast region, where the CME expansion begins early and is relatively fast. It is stretched out and quite distorted, reflecting the complex plasma dynamics in that region. Line 2 is southwest of the CME center, where the expansion begins later in time and is relatively slow. It shows a single large deflection from the CME and a smaller kink in the field line at the shock.

\begin{figure}%[h]
 \noindent\includegraphics[width=1.0\columnwidth]{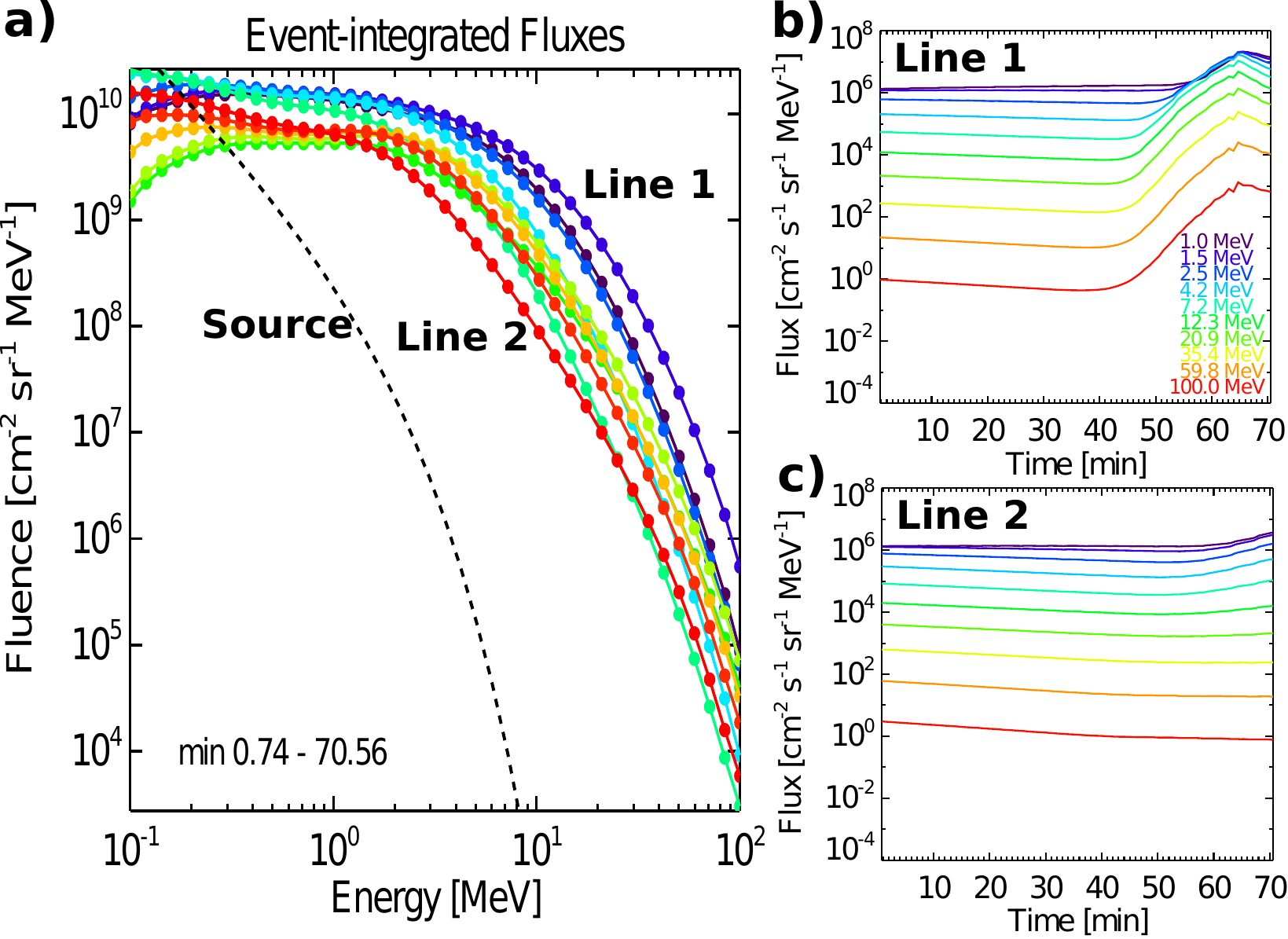}
 \caption{Panel a): event-integrated fluences for the 10 lines shown in Fig. \ref{figeprem_color_lines}. Proton energy is on the X-axis, fluence is on the Y-axis. The dashed line represents the fluence at the model inner boundary (source). Panels b) and c): simulated flux, `measured' over time at a constant distance of $\sim$8~\rsun on the two lines denoted in the bottom left panel of Fig. \ref{figeprem_color_lines} as Line 1 and Line 2, respectively. Time is on the X-axis, proton flux is on the Y-axis. }
 \label{figfluxes_fluences}
 \end{figure}

We inspect the fluxes and fluences `measured' at artificial observers located at $\sim$8~\rsun. Figure \ref{figfluxes_fluences}, panel a), shows the event-integrated fluxes (fluences) obtained from the pitch-angle-averaged proton distributions collected at $\sim$8~\rsun along the ten field lines shown in Fig. \ref{figeprem_color_lines}. Line 1 is in purple, Line 2 is in red. The two lines have the highest and lowest fluences, respectively, of all 10 lines, at most energies above 1~MeV. The Line 1 spectrum is harder than the Line 2 spectrum, and the overall fluences are also much higher, especially at energies higher than 5~MeV. The dashed line is the source spectrum.

Panels b) and c) show the flux time series on Lines 1 and 2. The time series in the two panels show the proton enhancements at the highest energy started around minute 40, when the shock front was near 3~\rsun (see Fig. \ref{figeprem_color_lines}). Peak fluxes were reached around minute 65 for Line 1, and around minute 70 for Line 2. More than three orders of magnitude enhancements at the highest energies (100~MeV) can be seen at the Line 1 observer, compared with no enhancement at that energy bin at the Line 2 observer.

We attribute the flux enhancements seen in the simulation to two factors. One is acceleration due to gradually changing plasma conditions along a given field lines. The second effect is more highly time-dependent acceleration due to the more sudden changes in the plasma along the field lines caused by the CME passage. The gradual enhancement gives the pre-CME suprathermal and energetic particle flux levels, which serve as seed populations for further diffusive acceleration after the CME is launched. Even though Line 2 is much more peripheral to the CME nose than Line 1, it was still influenced by the CME passage (see plasma plots in Fig.\ref{figl214}).

In order to gain better insight into what factors may contribute to the differences in proton acceleration along Lines 1 and 2,  we next examine concurrently the plasma parameters and relative flux enhancements at every node along those lines, for six snapshots of the simulation. We adopt a spatial dimension that is not along the radial direction, but instead runs along the field lines.

\begin{figure}%[h]
 \noindent\includegraphics[width=1.0\columnwidth]{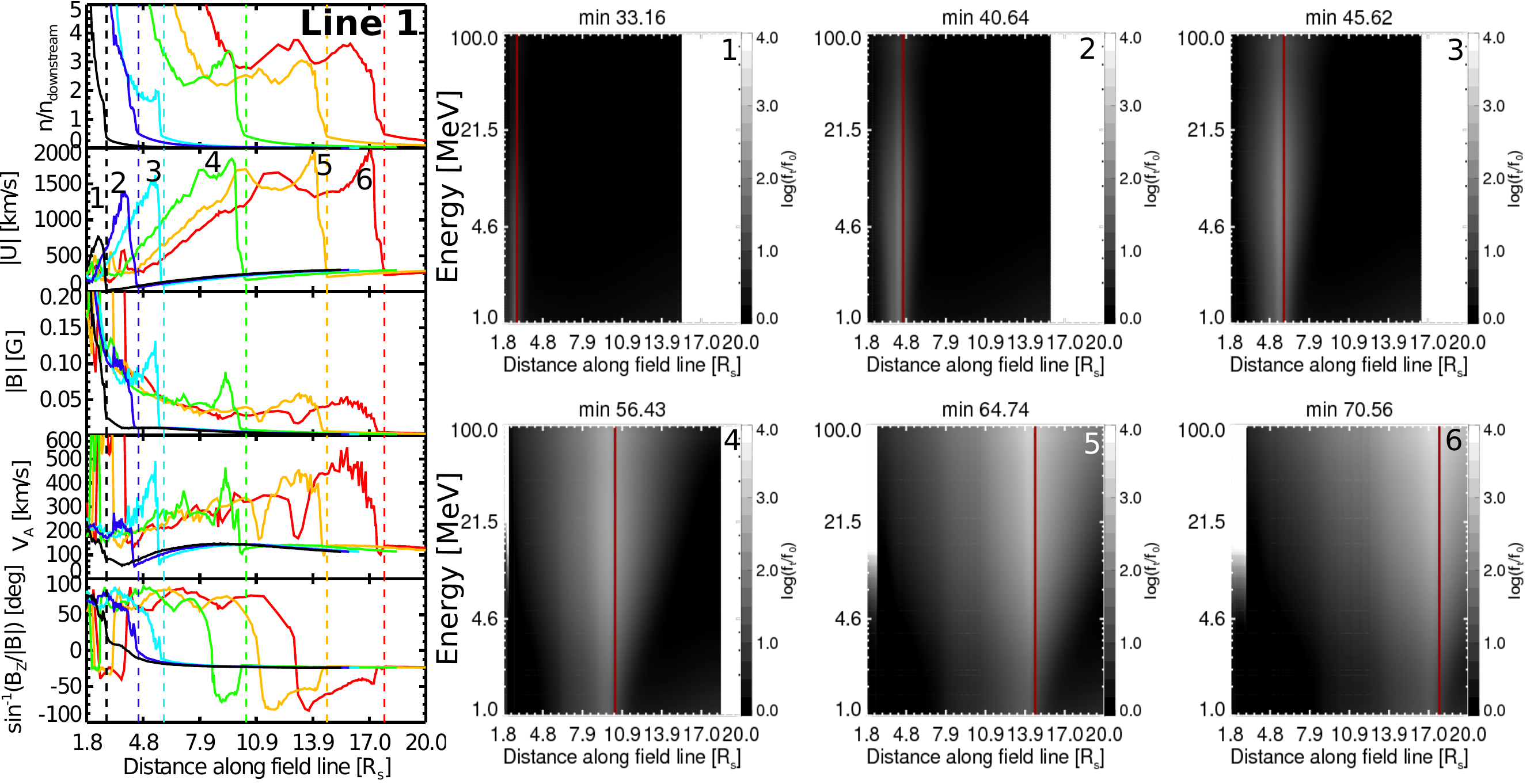}
 \caption{(Left side) Six MHD snapshots along Line 1 are shown in colors ranging from black to red, labeled 1-6. From top to bottom, the panels represent: number density relative to the downstream (see text), SW speed, magnetic field magnitude, Alfven speed, and angle $\theta_{Bz}$ (see text). The X-axis is distance along the field line. Vertical dashed lines denote the shock locations. (Right side) Contours of distribution enhancement for every snapshot. Distance along the line is on the X-axis, energy (1.0-100.0~MeV) is on the Y-axis. The grayscale represents $\log(f(t)/f(0))$, where f(t) is the distribution at every node and energy, and f(0) is the steady state distribution. Vertical red lines denote shock position.}
 \label{figl201}
 \end{figure}

\begin{figure}[h]
 \noindent\includegraphics[width=1.0\columnwidth]{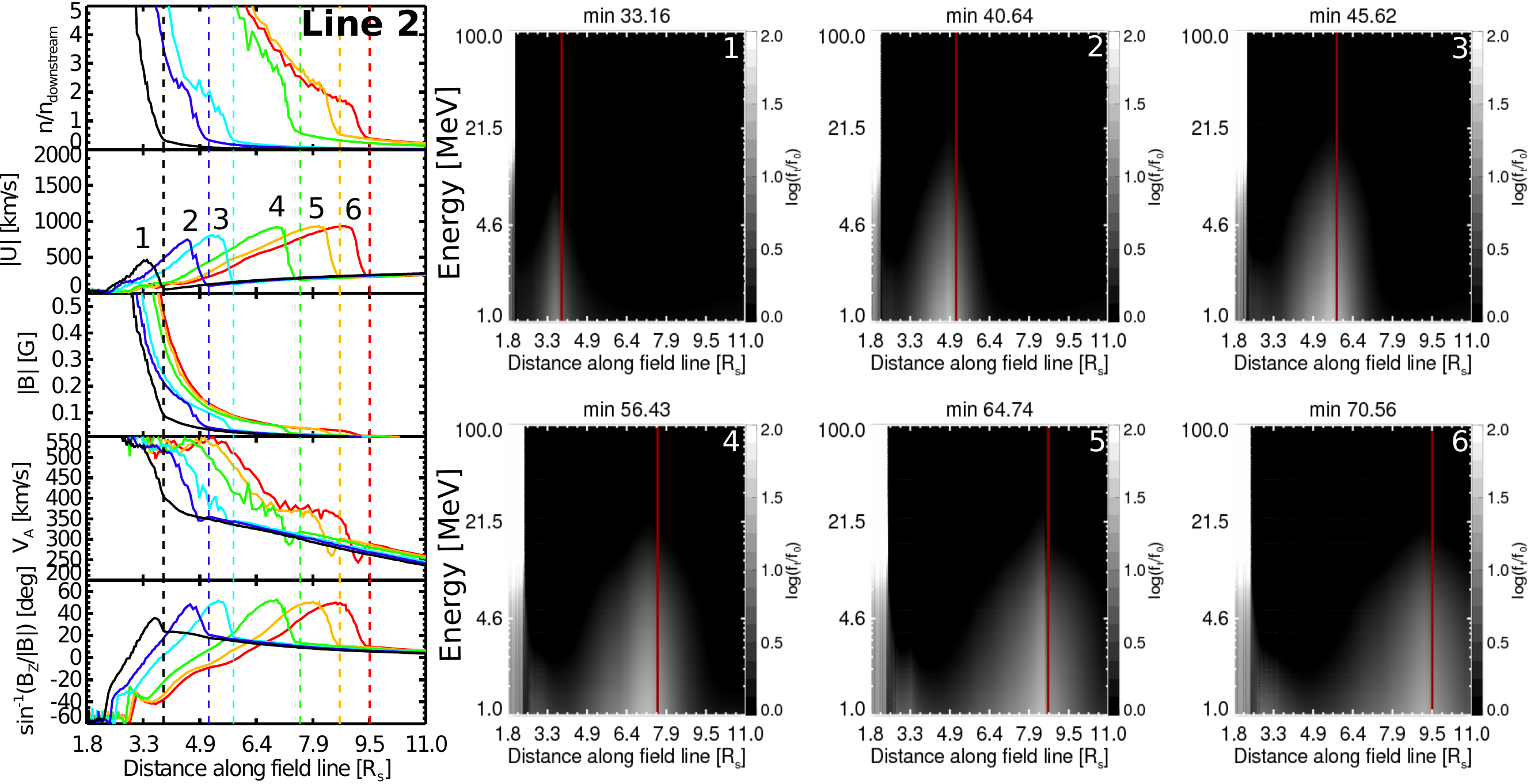}
 \caption{Same as Fig. \ref{figl201}, but for Line 2.}
 \label{figl214}
 \end{figure}

 %-------------------------------------------------SEP RESULTS---------------------------------------------------------------
 %\section{SEP Properties}
%\label{coronalresultsinner}
Figure \ref{figl201}, left side, shows the plasma conditions along Line 1, over the six consecutive snapshots similar in time as those in Fig. \ref{figdensity}. The five panels show, from top to bottom, density relative to the downstream, plasma speed, magnetic field magnitude, Alfven speed and angle $\theta_{Bz}$. The colored lines correspond to the six times from the simulation shown here, numbered in the middle panel. The top panel represents the ratio of the density along the field line to $n_{downstream}$, the average value of density over six points downstream of the instantaneous shock location. The angle $\theta_{Bz}=\sin^{-1}(B_Z/|\mathbf{B}|)$ quantifies how much of the magnetic field points in the Z-direction. It has been used previously by \citet{Evans:2011} in a slightly different form to separate regions of different magnetic field source: sharp changes in this angle mark boundaries between different structures. Here, the large gradient in this angle shows the interface between the CME flux rope and the solar wind field lines draped around it in the PUC region. The X-axis is distance along the field line, between 1.8 and 20~\rsun. The vertical dashed lines denote the position of the shock at every snapshot.

%EDIT THIS PARAGRAPH, ADD TO NEXT PARAGRAPH!!!
%{\bf EDIT!!!!}Observational and modeling studies of the Alfven speed profile in the corona \citep{Mann:2003, Evans:2008} show a characteristic shape of the profile marked by a local minimum of the speed between $\sim$1.3-3 \rsun followed by a local maximum and then a slow decline. This means that shocks may form low in the corona, then decay to simple waves where the Alfven speed is too high, and re-form as shocks beyond $\sim$6 R$_S$ \citep{Mann:2003}. This is supported by timing studies of energetic electrons, showing a delay between radio emission low in the corona and the release of electrons into interplanetary space\citep{Haggerty:2002}.
%EDIT THIS PARAGRAPH, ADD TO NEXT PARAGRAPH!!!

The shock front moves from left to right, followed by an intermediate speed enhancement in the sheath, and a maximum at the inner edge of the PUC, before the speed reduces gradually. The plots for the magnetic field magnitude and the density follow a similar shape, but the field enhancement is in general smoother. Although difficult, it is possible to separate two enhancements directly behind the shock - the first one in the sheath, the second one in the PUC. These are best seen in the last two snapshots (7-8), when the PUC has fully developed. The speed-up of the CME after it leaves the streamer is readily seen by comparing the shock front locations (vertical dashed lines) of time 1 with those at times 2-6 - the maximum speed jumps from 800 to 2000 km/s between steps 1 and 4. The Alfven speed closely follows the evolution of the magnetic field, peaking near the center of the PUC, as determined by the angle $\theta_{Bz}$ (in the bottom panel of the left side). This angle becomes abruptly negative behind the shock, close but not immediately adjacent to it, starting with step 3. In steps 4-6, the angle dip widens, possibly signifying the widening of the PUC, or at least an increase of the portion of the field line, which lies in the PUC.

In the right half of the figure there are six grayscale histograms, each corresponding to snapshots 1-6, and labeled accordingly. The X-axes have the same range as the left part of the figure, and proton energy in the range 1.0-100.0 MeV is on the Y-axes. The grayscale values represent the logarithm of the ratio of the (pitch angle-averaged) distribution function values at the particular snapshot at every node along the field line, and at every energy, to the distribution function of the steady state of the simulation before the CME erupts - $\log(f(x,E,t)/f(x,E,0))$. The dynamic range of the greyscale image is between 0.0 and 4.0, and the legends are shown to the right of every plot. The solid vertical red lines denote the shock position, just like on the left side of the figure.

In the first snapshot there is little relative enhancement, as this coincides with the CME being inside the streamer, expanding slowly. In the following steps an increase in the proton distribution enhancement can be seen. Several effects can be observed. First, the enhancement is initially greatest in the lower energies below 20 MeV, but then spreads out to higher energies (as may be expected), and remains largest at the highest energies, due to the fact that it is relative to the pre-CME distributions. Second, protons may be seen streaming away from the shock and sheath - the higher the energy, the farther they escape for a given time step. This causes the characteristic V-shape of the enhancement anti-sunward of the shock front. Finally, the maximum enhancement of the particle fluxes `travels' along the field line with the leading edge of the CME shock, as the protons are continuously accelerated. However, it is interesting to note that the maximum remains behind the shock, near the leading edge of the PUC. This could be a real effect, based on the additional, high compression and fast solar wind at the PUC. However, we must be cautious: MHD simulation shocks tend to be wider than real ones, which in the frame of shock acceleration means a reduction of the acceleration efficiency. While we are working to improve our MHD simulations, it is unclear whether the shock in this model would be more or less efficient than the PUC at accelerating protons.

The time-dependent plasma and SEP conditions along Line 2 are shown in Figure \ref{figl214}. The format is the same as the previous figure, as well as the time steps chosen. The range of distances along the field line plotted on both the left and right sides is 1.8-11~\rsun, since the CME is expanding more slowly in that region. The peak speed of the CME plasma along this line is less than half of the final speed along Line 1 ($\sim900$~km/s). The structure of the plasma parameters (left side) along the field line is less complicated and evolves similarly in time. The density only shows a single jump immediately behind the shock, in contrast with Line 1. The same is true for the plasma speed - a single jump behind the shock, followed by a slow decline. The magnetic field jump behind the shock decreases with time for this line, and does not show a secondary increase where the PUC would be. The last panel shows a slightly different behavior of the angle $\theta_{Bz}$ from that seen in the previous figure. It increases from zero to about 50~degrees, but then declines much more smoothly than seen in Line 1. This may be the signature of the shock sheath, which causes milder changes to the plasma behind the shock than a PUC region. In general, the front of the disturbance moves twice as slowly as along Line 1, and its speed does not increase significantly.

Looking at the right side of the figure (where the plots have a dynamic range between 0.0 and 2.0), a similar initial picture emerges as in Fig. \ref{figl201} - the proton enhancement occurs only at energies up to $\sim$20~MeV in the first time step. Here, however, there is no significant enhancement at the higher energies, and the relative enhancement is much weaker overall. The enhancement is again strongest closely behind the shock, due to the shock-sheath and not to a PUC. The location along Line 2 of the maximum enhancement is not as pronounced spatially as that in Line 1, perhaps due to the smoother changes in the MHD parameters behind the shock.

%DISCUSSION HERE
In general, protons along the lines in the northeastern region of the CME, where the expansion is fastest, have been accelerated much more than those in the southwestern region of slower expansion, especially at the high end of the energy spectrum. The lines on the northeastern side, where the CME exits the streamer first, show the greatest enhancements in the distribution function values, both in the time series and in the event-integrated fluences. The time to reach the peak of the profiles is about 10 minutes shorter along Line 1 than along Line 2. This is due to faster acceleration, as well as transport effects and the fact that the shock is moving much faster along Line 1. The acceleration timescale depends on the inverse square of the upstream speed, and is also proportional to the diffusion coefficient. The combination of these factors yields a shorter acceleration timescale along Line 1, compared with Line 2. The shock is also moving faster along Line 1, so the accelerated particles have a shorter distance to travel. There are four important results that our modeling work has pointed out. 1) Coronal magnetic field structure influences the CME shape and dynamics strongly, leading to different regions of the CME/shock developing different propagation speeds and compression regions. 2) These differences cause large changes to the amount of proton flux enhancement between field lines passing through different regions of the CME. 3) The largest particle fluxes occur immediately behind the shock wave (downstream) for all lines. 4) We observe additional major particle enhancements on field lines which pass through PUC regions, occurring where the lines cross the PUC leading edge.

The results point to the likely importance of the PUC region for acceleration near the Sun. The high speed of some parts of the CME causes a pile-up of material with a compression of the density and magnetic field comparable with or stronger than that at the shock, while in other parts there is no significant pile-up. Strong enhancement may occur due to the shock-like changes in the PUC plasma Ð big increases in density, magnetic field, and plasma velocity. This can occur in features that are not sharp transitions, and has been applied previously to explain acceleration at co-rotating interacting regions \citep{Giacalone:2002}. This mechanism, known as diffusive compression acceleration \citep{Giacalone:2005c}, is implicitly included in our numerical model. According to it, particles gain energy adiabatically in the compressive regions, and then diffuse away before the compensating cooling can occur. This scenario is likely also occurring in the PUC regions of CMEs. The enhancements at the PUC leading edge may thus be due to local acceleration, or a combination of acceleration and a pile-up effect. We will explore this effect in an upcoming publication.

One final point that we must discuss is the MHD representation of shocks. Due to computational and physical constraints, the shock is thicker in the MHD simulation than in reality, and spans a few computational cells. This is inevitable in a global simulation, which aims to cover a region of tens of solar radii, and execute in a finite amount of time (i.e., with a realistic grid resolution). Based on DSA theory, acceleration at a realistic, thinner shock would be more efficient. Thus, it is not entirely clear whether PUC acceleration is stronger in this simulation, since our estimates of acceleration at the shock are lower than in reality. However, we still note that the increase in velocity and the density compression are comparable at the sheath and at the shock (especially for the lines passing through the fast-expanding regions of the CME). Since the acceleration depends not only on the structure width, but also on changes to velocity, density, and magnetic field, it could still be possible to have stronger acceleration in the sheath. Future work will focus on improving shock resolution.
%-------------------------------------------------END SEP PROPERTIES---------------------------------------------------------------

 %--------------------------------------------CORONAL TRANSPORT SUMMARY--------------------------------------------------
 \subsection{Summary}
\label{coronalsummary}
%SUMMARY HERE
We present a detailed modeling study of global, CME-driven proton acceleration using a high-resolution coupling between results from a time-dependent global MHD simulation and a global kinetic acceleration and transport code. This framework will be used to characterize particle acceleration in the solar corona in a series of publications. Here, we study the May 13, 2005 CME as it has been studied before, and has well constrained CME dynamics. We modeled proton acceleration and transport along magnetic field lines in the corona as the CME evolved between distances of $\sim$1.8-8~\rsun. We first presented a brief analysis of the MHD simulation, in which we showed that the shape and dynamics of the CME are governed by the coronal conditions and the initial energy of the ejecta. Specifically, the CME evolution sped up significantly in regions where it exited the densest parts of the coronal streamer in which it was embedded. We found that proton acceleration is strongly influenced by the location of the field lines through which the CME passes. Our results show that efficient SEP acceleration could occur not only at shock fronts, but also in regions of strong CME expansion which compress the surrounding post-shock solar wind plasma. We quantified the differences in the time dependent fluxes and fluences in the lower to middle corona, and showed that significant acceleration does occur in this region. The results of this work may be applied to answering some of the major questions in space weather, for example whether CMEs in the corona may account for the acceleration of most fluxes in SEP events. Thus, we showed that CMEs in the low corona provide sites of strong acceleration, which may help explain early onsets of SEP events during CME initiation. 

%A more definitive way of determining the existence and extent of PUC regions along field lines will be explored in an upcoming study. 
%The next step is to continue the simulations out to 1 AU and beyond, to compare with the plethora of data in the heliosphere. 

 %----------------------------------------END CORONAL TRANSPORT SUMMARY-----------------------------------------------

\section*{Acknowledgements}
This work was supported under NASA LWS EMMREM project and grant no. NNX07AC14G. K.A.K. was partially supported under the NASA Living With a Star Jack Eddy Postdoctoral Fellowship Program, administered by the UCAR Visiting Scientist Programs. R. M. E. is supported through an appointment to the NASA Postdoctoral Program at GSFC, administered by Oak Ridge Associated Universities through a contract with NASA.

%\addcontentsline{toc}{section}{References}

%\newpage
%\bibliography{coronal_acceleration_paper}

\end{document}